\documentclass[lettersize,journal]{IEEEtran}
\usepackage{amsmath,amsfonts}
\usepackage{algorithm}
\usepackage[noend]{algpseudocode}
\usepackage{etoolbox}

\usepackage{array}
\usepackage[caption=false,font=normalsize,labelfont=sf,textfont=sf]{subfig}
\usepackage{textcomp}
\usepackage{stfloats}
\usepackage{url}
\usepackage{verbatim}
\usepackage{lipsum}
\usepackage{graphicx}
\usepackage{epstopdf}

\usepackage{cite}

\usepackage{geometry}

\makeatletter
\newenvironment{breakablealgorithm}
  {% \begin{breakablealgorithm}
   \begin{center}
     \refstepcounter{algorithm}% New algorithm
     \hrule height.8pt depth0pt \kern2pt% \@fs@pre for \@fs@ruled
     \renewcommand{\caption}[2][\relax]{% Make a new \caption
       {\raggedright\textbf{\ALG@name~\thealgorithm} ##2\par}%
       \ifx\relax##1\relax % #1 is \relax
         \addcontentsline{loa}{algorithm}{\protect\numberline{\thealgorithm}##2}%
       \else % #1 is not \relax
         \addcontentsline{loa}{algorithm}{\protect\numberline{\thealgorithm}##1}%
       \fi
       \kern2pt\hrule\kern2pt
     }
  }{% \end{breakablealgorithm}
     \kern2pt\hrule\relax% \@fs@post for \@fs@ruled
   \end{center}
  }

\def\therule{\makebox[\algorithmicindent][l]{\hspace*{.5em}\vrule height .75\baselineskip depth .25\baselineskip}}%

\newtoks\therules% Contains rules
\therules={}% Start with empty token list
\def\appendto#1#2{\expandafter#1\expandafter{\the#1#2}}% Append to token list
\def\gobblefirst#1{% Remove (first) from token list
  #1\expandafter\expandafter\expandafter{\expandafter\@gobble\the#1}}%
\def\LState{\State\unskip\the\therules}% New line-state
\def\pushindent{\appendto\therules\therule}%
\def\popindent{\gobblefirst\therules}%
\def\printindent{\unskip\the\therules}%
\def\printandpush{\printindent\pushindent}%
\def\popandprint{\popindent\printindent}%

\algdef{SE}[WHILE]{While}{EndWhile}[1]
  {\printandpush\algorithmicwhile\ #1\ \algorithmicdo}
  {\popandprint\algorithmicend\ \algorithmicwhile}%
\algdef{SE}[FOR]{For}{EndFor}[1]
  {\printandpush\algorithmicfor\ #1\ \algorithmicdo}
  {\popandprint\algorithmicend\ \algorithmicfor}%
\algdef{S}[FOR]{ForAll}[1]
  {\printindent\algorithmicforall\ #1\ \algorithmicdo}%
\algdef{SE}[LOOP]{Loop}{EndLoop}
  {\printandpush\algorithmicloop}
  {\popandprint\algorithmicend\ \algorithmicloop}%
\algdef{SE}[REPEAT]{Repeat}{Until}
  {\printandpush\algorithmicrepeat}[1]
  {\popandprint\algorithmicuntil\ #1}%
\algdef{SE}[IF]{If}{EndIf}[1]
  {\printandpush\algorithmicif\ #1\ \algorithmicthen}
  {\popandprint\algorithmicend\ \algorithmicif}%
\algdef{C}[IF]{IF}{ElsIf}[1]
  {\popandprint\pushindent\algorithmicelse\ \algorithmicif\ #1\ \algorithmicthen}%
\algdef{Ce}[ELSE]{IF}{Else}{EndIf}
  {\popandprint\pushindent\algorithmicelse}%
\algdef{SE}[PROCEDURE]{Procedure}{EndProcedure}[2]
   {\printandpush\algorithmicprocedure\ \textproc{#1}\ifthenelse{\equal{#2}{}}{}{(#2)}}%
   {\popandprint\algorithmicend\ \algorithmicprocedure}%
\algdef{SE}[FUNCTION]{Function}{EndFunction}[2]
   {\printandpush\algorithmicfunction\ \textproc{#1}\ifthenelse{\equal{#2}{}}{}{(#2)}}%
   {\popandprint\algorithmicend\ \algorithmicfunction}%
\makeatother
\begin{document}

\title{RIS-assisted Seamless Connectivity in Wireless Multi-Hop Relay Networks}

\author{Peini Yi,~\IEEEmembership{Student Member,~IEEE}, Wenchi Cheng,~\IEEEmembership{Senior Member,~IEEE}, Jingqing Wang,~\IEEEmembership{Member,~IEEE}\\ and Wei Zhang,~\IEEEmembership{Fellow,~IEEE} 
        % <-this % stops a space
\thanks{Peini Yi, Wenchi Cheng and Jingqing Wang are with State Key Laboratory of Integrated Services Networks, Xidian University, Xian 710071, China (e-mails: pnyi@stu.xidian.edu.cn; wccheng@xidian.edu.cn; jqwangxd@xidian.edu.cn.)}% <-this % stops a space
\thanks{Wei Zhang is with the School of Electrical Engineering and Telecommunications, University of New South Wales, Sydney, NSW 2052, Australia
(e-mails: w.zhang@unsw.edu.au)}}

%\markboth{IEEE JOURNAL ON SELECTED AREAS IN COMMUNICATIONS,~Vol.~, No.~, May~2024}%

\IEEEpubid{}
% Remember, if you use this, you must call \IEEEpubidadjcol in the second

\maketitle
\thispagestyle{empty}

\begin{abstract}
    In recent years, reconfigurable intelligent surfaces (RIS) have garnered significant attention for their ability to control the phase shifts in reflected signals. By intelligently adjusting these phases, RIS can establish seamless direct paths between communication devices obstructed by obstacles, eliminating the need for forwarding and significantly reducing system overhead associated with relaying. This capability is crucial in multi-hop ad hoc networks requiring multiple relay steps. Consequently, the concept of incorporating multi-hop RIS into wireless multi-hop relay networks has emerged.
    In this paper, we propose a novel network model where each UAV communication node is equipped with a RIS, facilitating seamless connections in multi-hop relay wireless networks. We analyze the performance of this model by integrating RIS-assisted physical layer modeling into the seamless connection network framework and conducting a detailed comparative analysis of RIS-assisted and conventional connections. At the medium access layer, we introduce a RIS-DCF MAC protocol based on the IEEE 802.11 distributed coordination function (DCF), modeling the medium access process as a two-hop access scenario. Our results demonstrate that the seamless connections and diversity gain provided by RIS significantly enhance the performance of multi-hop relay wireless networks.

\end{abstract}

\begin{IEEEkeywords}
    Reconﬁgurable intelligent surface (RIS), 802.11, MAC protocol, Mobile ad hoc network (MANET).
\end{IEEEkeywords}

\section{Introduction}
\IEEEPARstart{R}{econfigurable Intelligent Surface }(RIS) has emerged as an up-and-coming technology, recognized for its potential to significantly enhance coverage range and boost communication performance in future wireless networks beyond 5G and 6G \cite{chenReconfigurableintelligentsurfaceassistedB5G6G2023}. This innovation has garnered extensive attention from both academia and industry. Unlike traditional relays, RIS relays do not require any form of forwarding. By intelligently modifying the phase shift of the reflected signal, RIS can create a seamless direct path and achieve diversity gains between two communication devices, even in obstructed environments \cite{yildirimHybridRISEmpoweredReflection2021,xuReconfiguringWirelessEnvironments2023}. This capability greatly reduces the transmission delay and system overhead typically generated by traditional relays. Therefore, RIS has broad application prospects in ad hoc networks involving massive multi-hop data forwarding.

The concept of ad hoc networks has been around for over half a century since its inception. As a self-organizing multi-hop wireless communication network, it has continuously developed and expanded in both technology and application \cite{chakeresAODVRoutingProtocol2004,ramanathanAdHocNetworking2005} fields since its initial conception and implementation. Originally designed to meet the requirements of military communication, ad hoc networks have gradually evolved and diversified, forming various categories to adapt to different application scenarios and requirements. The Mobile Ad hoc Network (MANET), as the core concept of ad hoc networks, provides the foundation for specific application networks such as the Flight Ad hoc Network (FANET) and the Vehicle Ad hoc Network (VANET) \cite{contiMultihopAdHoc2007}. These specific networks have been optimized and expanded based on MANET, targeting their unique application environments and node characteristics. FANET applies the concept of ad hoc networks to unmanned aerial vehicular (UAV) swarms \cite{mansoorFreshLookRouting2023,sariChainRTSCTS2021}. At the same time, VANETs are a crucial part of the vehicle to everything (V2X) \cite{bhoverV2XCommunicationProtocol2017}, applying the ad hoc concept to vehicle communication to support intelligent transportation systems. 

Considering mobility and payload capacity attributes, vehicles emerge as highly conducive platforms for integrating RIS technology. This is especially relevant for UAVs, owing to significant improvements in drone control precision \cite{czyzaAssessmentAccuracyUnmanned2023}, a topic of extensive research and empirical validation \cite{mursiaRISeFlightRISempowered2021, yaoResourceAllocation5GUAVBased2021b, chenReconfigurableIntelligentSurface2023,ataRISEmbeddedUAVsCommunications2024}. The deployment of aerial RIS (ARIS) has been instrumental in extending the reach of fundamental communication infrastructure. This advancement is especially beneficial in emergency scenarios where conventional communication infrastructure has been compromised. Nonetheless, the current research predominantly concentrates on deploying UAVs equipped with RIS modules, or ARIS, to enhance communication between terrestrial user equipment and cellular base stations. In stark contrast, the potential utility of RIS-aided communication among UAV swarms has not been adequately investigated, representing an untapped area for future research and development.

Multi-hop RIS-assisted wireless multi-hop relay networks involve physical layer design and media access control (MAC) layer design. On the one hand, the existing research on multi-hop RIS-assisted physical layer is very comprehensive, mainly focusing on beamforming \cite{guoJointCommunicationSensing2023,huangMultiHopRISEmpoweredTerahertz2021} and performance analysis \cite{chapalaReconfigurableIntelligentSurface2022,chapalaMultipleRISAssistedMixed2023}. Significantly, in the study by Chapala et al. \cite{chapalaReconfigurableIntelligentSurface2022}, the authors not only provided a point-to-point communication model and performance details for multi-hop cascaded channels in RIS but also highlighted that as the number of hops in cascading multi-hop RIS increases, it will decrease the diversity gains introduced by RIS. 
Additionally, several studies have investigated the performance comparison between RIS-based and traditional relays \cite{abdullahDoubleRISCommunicationDF2022,zhouOperatorPerspectiveHow2023}. In particular, Zhou et al. \cite{zhouOperatorPerspectiveHow2023} analyzed RIS performance in amplify-and-forward (AF) and decode-and-forward (DF) modes under both full- and half-duplex scenarios. Their findings demonstrated that RIS eliminates the need for direct signal reception at the relay, thereby enhancing physical layer communication performance and emphasizing the importance of RIS's seamless relay capabilities.

On the other hand, most existing research on RIS-assisted MAC layer protocols focuses on multiple users accessing networks through RIS modules independent of communication devices \cite{caoReconfigurableIntelligentSurfaceAssistedMACWireless2021,caoMassiveAccessStatic2022,zhangDistributedCSMACA2023}. In addition, there have been some recent studies on RIS-assisted MANET. Phung et al. \cite{phungMaximizingThroughputRouting2024} proposed an indoor RIS-assisted multi-hop mesh communication scheme, providing an analysis model for the interaction between user devices, relay devices, and the RIS module. 
In \cite{tranQoSMulticastRouting2024}, Tran et al. provided a QoS multicast routing protocol for RIS-assisted MANET. All the RIS-assisted modules in this article are connected to the central processing unit with DNN through a control link for coordination with the base station. The RIS modules in these two articles assist MANET through centralized control rather than distribution. In summary, there is a notable gap in research concerning the integration of RIS with communication equipment and the collaborative operation between RIS and such equipment.

Unlike various existing studies on the application of RIS in the MAC layer, our research distinguishes itself by integrating the RIS module with the communication system of multi-hop relay network nodes, where each node is equipped with its own RIS. This approach fully exploits the diversity gain and seamless relay capability of multi-hop RIS, enabling distributed and independent RIS usage between nodes. By allowing nodes to autonomously negotiate RIS usage, our work offers a novel solution to ad-hoc relay methods and enhances MAC layer networking performance, particularly in dynamic, decentralized environments.
However, this approach presents several challenges. First, a feasible network model framework must be proposed to effectively integrate multi-hop RIS into the multi-hop relay network and enable their coordinated operation. Second, to explore the performance improvements of RIS as a physical layer device for distributed multi-hop node networking, it is essential to establish a physical channel model to quantify its diversity gain. Third, to evaluate the benefits of multi-hop seamless connections, developing an essential and effective RIS-assisted multi-hop MAC layer protocol for studying the fundamental protocol performance while avoiding the hidden terminal problems the degradation caused by multi-hop cascaded RIS is necessary.

In this paper, we aim to model and analyze a seamless connection in multi-hop RIS-assisted wireless multi-hop relay networks. First, we propose a general network model where each node is equipped with a RIS to provide a seamless link between the source and destination nodes. We design a protocol based on the 802.11 distributed coordination function (DCF) called RIS-DCF. To develop and analyze the effectiveness of the RIS-DCF protocol, we propose a multi-hop seamless connection MAC protocol and address the issue of RIS channel reservation in wireless networks. Additionally, we aim to mitigate hidden terminal problems within the network. Furthermore, to evaluate the RIS-DCF MAC protocol, we establish an analytical model to derive the system throughput of the RIS-DCF network.

The rest of this paper is organized as follows. 
Section II presents the network system model and the physical layer transmission model and defines RIS channel utilization efficiency. 
Section III introduces our proposed RIS-DCF protocol and explains its use in the network system. 
Section IV proposes a dual-hop access collision model for all RIS-assisted and conventional channel models under our proposed RIS-DCF protocol.
Section V conducts numerical analyses to evaluate our proposed RIS-DCF protocol and compare its performances with those when using convention half-duplex MAC protocols. 
The paper concludes with Section VI.

\begin{figure*}[t]
    \centering
    \includegraphics[width=6.7in]{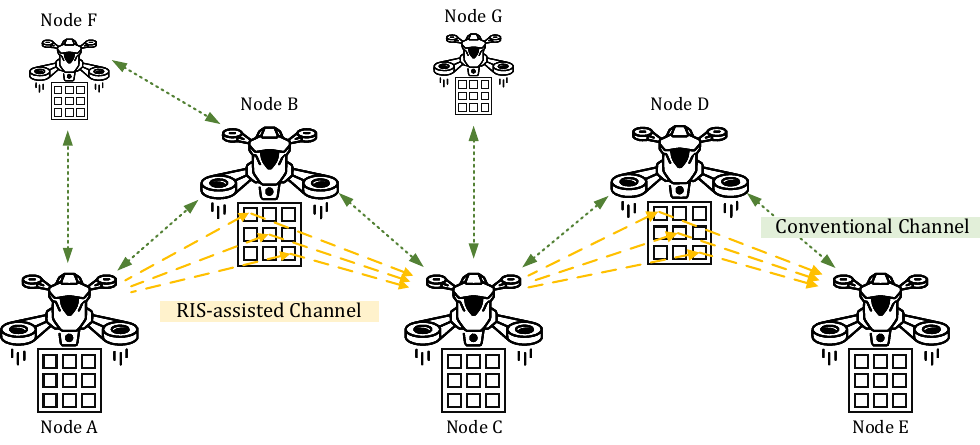}
    \caption{RIS-Assisted seamless link communication network scenario.}
    \label{fig_1}
  \end{figure*}
\section{System Model}

In this section, we propose a MANET network model where each UAV carries a RIS. The RIS-assisted channel is integrated into our multi-hop relay network at the physical layer to characterize the access capability gain of the RIS-assisted network system.

\subsection{Network Model}

We consider a RIS-assisted multi-user decentralized wireless network, where each UAV node is equipped with a RIS controller that can be operated independently, and the RIS module consists of \( N \) passive elements.
The model is illustrated in Fig.~1. Each UAV node maintains a hovering state throughout the transmission cycle and can operate as a source node, relay node, or destination node, depending on its role in the communication process. The RIS modules enhance the UAV network by utilizing the original air interface of the UAV nodes.

When the RIS is inactive, all nodes in the network are connected to each other through conventional channel links, indicated by the green dashed lines in Fig.~1. For instance, Node D is connected to Node E through a conventional channel. At this point, we assume that nodes unable to detect each other lack a line-of-sight (LOS) channel and cannot establish a connection due to obstructions. In such scenarios, based on the existing ad hoc architecture \cite{ramanathanBriefOverviewAd2002, carranoIEEE80211s2011}, two non-adjacent nodes can communicate through multi-hop relays.

When the RIS is active, the communication process avoids using a cascaded multi-hop RIS channel to prevent performance degradation \cite{chapalaReconfigurableIntelligentSurface2022}. As shown in Fig.~1, if Node A needs to send data packets to Node E, which is four hops away, the process is as follows: First, data packets are relayed from Node A to Node C through the RIS module of Node B. Node C, acting as a relay node, then forwards the data packets to Node E through the RIS module of Node D.

Notably, conventional channels are still used to transmit data packets even when RIS-assisted links are employed. Consider a scenario illustrated in Fig.~1, where Node A transmits packets to Node G, which is three hops away. The process begins with Node A relaying a packet to Node C via the RIS module of Node B. Subsequently, Node C, acting as a relay node, transmits the packet to Node G over conventional channels. Conventional ad hoc network protocols typically divide multi-hop routes into a sequence of single-hop connections. In contrast, the proposed RIS-assisted network architecture decomposes multi-hop routes into dual-hop and single-hop connections, enhancing diversity gain while mitigating the complexity of multi-hop channel state information (CSI) estimation. This architectural distinction significantly impacts the design and analysis of MAC protocols.

\subsection{Channel Model}

In the wireless system illustrated in Fig.~1, two types of physical layer channel models are utilized: the conventional link channel and the RIS-assisted link channel. The conventional link channel represents the LOS channel between neighboring nodes. In contrast, the RIS-assisted link channel enables non-adjacent communication between dual-hop nodes facilitated by the RIS. The subsequent sections provide an analysis of these two channel models.

First, we consider the conventional link channel between Node D and Node E, where Node D is the source node and Node E is the destination node. Next, we examine the RIS-assisted link channel scenario, where Node A is the source node, Node B is the relay node, and Node C is the destination node. It is important to note that there is no direct link between Node A and Node C, and the communication from Node A to Node C occurs solely through the RIS channel controlled by Node B.

Based on the above model, the received signal \( y_{DE}^{C} \) from source node D to the destination node E in the conventional channel can be expressed as:
\begin{equation}
y_{DE}^{C}(t) = \sqrt{P_s}H_{DE}s_{D}(t) + w(t)
\label{eql_1}
\end{equation}
where \( H_{DE} \) is the channel from source node to destination node, \( s_D \) is the transmitted signal from source node, and \( w(t) \) is modeled as \( \mathcal{CN}(0, N_0) \) Gaussian white noise.

Similarly, we can derive the received signal \( y_{AC}^{R}(t) \) for the RIS-assisted channel model from source node A to destination node C through relay node B's RIS assistance:
\begin{equation}
y_{AC}^{R}(t) = \sqrt{P_s}\mathbf{H}_{AB} \boldsymbol{\Theta}_B \mathbf{H}_{BC} s_S(t) + w(t)
\label{eql_2}
\end{equation}
where the set \( \mathbf{H}_{AB} = \{H_{AB}^1, H_{AB}^2, \ldots, H_{AB}^N\} \) and the set \( \mathbf{H}_{BC} = \{H_{BC}^1, H_{BC}^2, \ldots, H_{BC}^N\} \) represent the channels from node A to node B and from B to C under node B's RIS assistance, respectively, divided into \( N \) direct paths according to the number of RIS elements. \( w(t) \) is modeled as \( \mathcal{CN}(0, N_0) \) complex Gaussian white noise, and the set \( \boldsymbol{\Theta}_B = \operatorname{diag}(e^{j \theta^1}, \ldots, e^{j \theta^N}) \) is the phase shift matrix for the \( N \)-RIS elements, with \( e^{j \theta^n} \) being the phase shift for the \( n \)-th RIS element, where \( n \leq N \).

For the conventional channel \( H_{DE} \) in Eq. \eqref{eql_1} and the RIS-assisted channel \( H_{AB}^n \), \( H_{BC}^n \) in Eq. \eqref{eql_2}, the detailed expressions are given below:
\begin{equation}
    \left\{
    \begin{array}{ll}
 H_{DE} &= \sqrt{L_{DE}} \left|\tilde{h}_{DE}\right| e^{j \alpha} , \\
 H_{AB}^n &= \sqrt{L_{AB, n}} \left|\tilde{h}_{AB, n}\right| e^{j \phi_n} , \\
 H_{BC}^n &= \sqrt{L_{BC, n}} \left|\tilde{h}_{BC, n}\right| e^{j \omega_n}        
    \end{array}
    \right.
\end{equation}
where \( \alpha \), \( \phi_n \), and \( \omega_n \) represent the phases of the LOS signal from Node D to Node E in the conventional channel, from Node A to the \( n \)-th element of the RIS carried by Node B, and from the \( n \)-th element of the RIS carried by Node B to Node C in the RIS-assisted channel, respectively. Similarly, \( \sqrt{L_{DE}} \), \( \sqrt{L_{AB, n}} \), and \( \sqrt{L_{BC, n}} \) represent the path losses for the respective channels based on the Friis free space propagation model, where $ \sqrt{L(\cdot)} = \frac { \lambda } { 4 \pi d(\cdot)}$. The terms \( \left| \tilde{h}_{DE} \right| \), \( \left| \tilde{h}_{AB, n} \right| \), and \( \left| \tilde{h}_{BC, n} \right| \) denote the channel fading for the respective channels, following the Nakagami-m distribution.

\subsection{RIS Efficiency}

To evaluate the link cost and diversity gain of the RIS-assisted system, we introduce the utilization efficiency of the RIS channel relative to the conventional LOS channel, denoted by $\eta$. This metric represents the ratio of the transmission rate achieved over the RIS-assisted channel to that of the conventional LOS channel. Specifically, $\eta$ quantifies the improvement in transmission efficiency due to the integration of RIS. The utilization efficiency $\eta$ is mathematically expressed as follows:

\begin{equation}
\label{eta}
\eta = \frac
{\mathbb{E} [\log_2(1+\gamma_{R})]}
{\mathbb{E} [\log_2(1+\gamma_{C})]} 
\end{equation}
where $\gamma_{R}$ is the signal-to-noise ratio (SNR) in the RIS-assisted channel scenario, and $\gamma_{C}$ is the SNR in the conventional channel scenario.
$\mathbb{E}[\log_2(1+\gamma)]$ is the ergodic achievable rate in both scenario. For the RIS-assisted channel scenario, we refer to \cite{niPerformanceAnalysisRISassisted2021} to calculate our results. For the conventional channel scenario, we refer to \cite{goldsmithWirelessCommunications} for our calculations.

Next, we will analyze the received SNR.
According to the definition of received SNR and the formula \cite{goldsmithWirelessCommunications}, we can give the SNR $\gamma_{C}$ for the conventional channel as follows:
\begin{equation}
\gamma_{C}= \frac{P_s}{N_0}| H_{DE}|^2 = \frac{P_s}{N_0}\left( \frac { \lambda } {4 \pi d } \right) ^ { 2 }\left| \tilde{h}_{SD}\right|^2
\end{equation}
where $\lambda$ is the wavelength, and $d$ is the distance between node S and node D. 
Similarly, the received SNR $\gamma_{R}$ for the RIS-assisted system can be expressed as:
\begin{equation}
\label{deqn_2b3}
\gamma_{R}= 
\frac{P_s}{N_0}
\left( \frac { \lambda } { 4 \pi } \right) ^ { 4 } 
\left| \sum _ { n = 1 } ^ { N } \frac{ \left| \tilde{h}_{SR,n} \right| \left|\tilde{h}_{RD,n}\right| e^{j\left(\phi_n+\omega_n+\theta_n\right)}}{d_{SR,n}d_{RD,n}} \right| ^ { 2 }
\end{equation}
where $d_{SR,n}$ and $d_{RD,n}$ are the distances from node S to the $n$-th element of the RIS at node R and from the $n$-th element of the RIS at node R to the $n$-th element at node D, respectively.

Assuming that the phase shifts on the relay node R have been intelligently adjusted through phase matching to achieve optimal performance, in this case, all CSI is fully known at the receiving node through various channel estimation methods \cite{xuDeepLearningBased2021,yuanReconfigurableIntelligentSurfaceEmpoweredWireless2021}. Also, we assume the distances between all nodes are equal, i.e., $d = d_{SR} = d_{RD} \gg N\lambda$. Considering this ideal scenario, $\gamma_{R}$ in Eq. \eqref{deqn_2b3} can be simplified to:
\begin{equation}
    \gamma_{R}= \frac{P_s}{N_0}
    \left( \frac { \lambda } { 4 \pi d} \right) ^ { 4 } 
    \left| \sum _ { n = 1 } ^ { N } {\left|\tilde{h}_{SR}^n \right| \left|\tilde{h}_{RD}^n\right|} \right| ^ { 2 }
\end{equation}

\section{RIS DCF-Based MAC Protocal}

\begin{figure*}[t]
    \centering
    \includegraphics[width=7in]{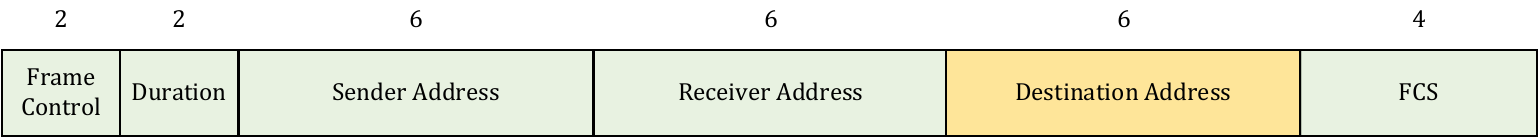}
    \caption{R-RTS/R-CTS headers in RIS-DCF protocol.}
    \label{fig_2h}
  \end{figure*}

Once the network model has been established, it is necessary to propose a MAC protocol to demonstrate its functionality \cite{chengFullDuplexSpectrumSensingMACProtocol2015,chengRTSFCTSMechanism2013}. We have developed our RIS-DCF protocol, which builds upon the conventional Request-to-Send (RTS)/Clear-to-Send (CTS) mechanism \cite{bianchiPerformanceAnalysisIEEE2000}. 
In this section, we first introduce our RIS-assisted DCF (RIS-DCF) protocol. Subsequently, we comprehensively explain how our RIS-DCF protocol can be applied in RIS-assisted networks. Finally, we demonstrate the cooperation of other nodes within the network and address how to tackle the problem of hidden nodes.

\subsection{RIS-DCF MAC Protocol Overview}

In a RIS-assisted network, each node is equipped with a RIS module and controller to perform the node's relay tasks. We choose to develop the protocol based on distributed access rather than centralized access to enhance the system's flexibility, scalability, and fault tolerance, while reducing latency and simplifying the network architecture to better adapt to dynamically changing network environments.

Before developing the RIS-DCF protocol, it is necessary to examine whether the MAC protocol relies on the RTS/CTS or ACK mechanisms. The ACK-based mechanism is more straightforward and does not necessitate channel reservation, yet it confronts the issue of hidden terminals \cite{xuEffectivenessRTSCTS2003}. The RTS/CTS mechanism can address the corresponding issue of hidden terminals. Beyond the issue of hidden terminals, the RIS-DCF protocol necessitates reservations for using RIS channels. The relay node R, which employs the RIS to forward messages, must estimate and calculate the phases of each RIS element for the channel from the optimized source node s to the destination node D through a handshake process. Thus, we turn to develop the RTS/CTS-based RIS-DCF protocol for RIS-aided networks. 

In our RIS-DCF protocol, we utilize RTS and CTS frames to complete the handshake process. To enable the RTS frame to identify the destination node, we have extended the RTS/CTS header from the original IEEE 802.11 protocol\cite{IEEEStandardInformation2011}. As illustrated in Fig.~2, we add a destination node address field to the header and name the new frames RIS-assisted RTS (R-RTS) and RIS-assisted CTS (R-CTS). This additional field is crucial for selecting the next hop to the destination or determining whether the current node is the final destination.

We can divide nodes using different strategies in the network into two types. One type is the first node that sends R-RTS, which we call the sending node and denote as $\mathbf{X}$. Another type is the node that receives R-RTS, which we call the receiving node, denoted as $\mathbf{Y}$. For any node in this network, both the sending and receiving strategies must be supported, and their difference lies only in the different triggering methods. The routing methodology in RIS-DCF protocol follows traditional MANET protocols, such as DSDV (Destination-Sequenced Distance Vector) and AODV (Ad-hoc On-demand Distance Vector), with the routing table assumed to be pre-established.

We then describe our proposed RIS-DCF protocol in pseudocode as follows.

\begin{breakablealgorithm}
    \caption{RIS-DCF Protocol for Sending Nodes $\mathbf{X}$}
    \begin{algorithmic}[1]
        \State $\mathbf{X}$ has data to transmit.
        \State According to the routing table, $\mathbf{X}$ finds the next hop of the data destination node and sends R-RTS to the next hop.
        \If{collision detected}
            \LState Wait for random backoff time.
            \LState Send R-RTS again.
        \Else
            \LState $\mathbf{X}$ waits for a SIFS time for R-CTS or R-RTS.
            \If{no R-RTS/R-CTS received}
                \LState Resend R-RTS request to the next hop.
            \ElsIf{R-RTS with $\mathbf{X}$ as the source received}
                \LState $\mathbf{X}$ waits for a SIFS time for R-CTS.
            \ElsIf{R-CTS received}
                \If{R-RTS's destination $\neq$ the next hop}
                    \LState Extract the R-CTS Sender Address.
                    \LState $\mathbf{X}$ sends data to the R-CTS Sender.
                \ElsIf{R-RTS's destination $=$ the next hop}
                    \LState $\mathbf{X}$ sends data to the next hop.
                \EndIf
                \LState $\mathbf{X}$ waits for a SIFS time for ACK.
                \If{no ACK received}
                    \LState Increment retry counter.
                    \If{retry counter exceeds retry limit}
                        \LState Transfer failed. Abort transmission.
                    \Else
                        \LState Apply backoff before retrying.
                        \LState Transfer failed. Try again after backoff.
                    \EndIf
                \EndIf
            \EndIf
        \EndIf
    \end{algorithmic}
\end{breakablealgorithm}

The RIS-DCF protocol for sending nodes optimizes data transmission by selecting the appropriate next hop, handling collisions through a backoff mechanism and a retransmission mechanism, and implementing retry limits with exponential backoff to prevent network congestion. This protocol prioritizes data integrity and transmission reliability while ensuring optimal network performance. Additionally, it dynamically adjusts its parameters based on network conditions, such as node density and traffic load, to maintain high throughput and low latency. The adaptive nature of this protocol makes it robust in diverse networking environments, contributing significantly to the overall stability and efficiency of the communication system. 

\begin{breakablealgorithm}
    \caption{RIS-DCF Protocol for Receiving Nodes $\mathbf{Y}$}
    \begin{algorithmic}[1]
        \State $\mathbf{Y}$ received an R-RTS from the previous hop node.
        \If{the destination of R-RTS is $\mathbf{Y}$}
            \LState $\mathbf{Y}$ replies R-CTS.
            \LState $\mathbf{Y}$ waits for a SIFS time for DATA.
            \If{$\mathbf{Y}$ received DATA from the source node}
                \LState Reply ACK to the previous hop node.
            \Else
                \LState Abort current operation.
                \LState Enter backoff state, then send R-CTS again.
            \EndIf
        \Else
            \If{RIS module of $\mathbf{Y}$ operates normally}
                \LState Transfer R-RTS to the next hop.
                \LState Adjust the RIS module phase for RIS link.
                \If{collision detected}
                \LState Enter backoff state, then send R-RTS again.
                \Else
                    \LState Wait for R-CTS from the destination.
                    \If{$\mathbf{Y}$ receives R-CTS from next hop}
                        \LState Extract duration of the R-CTS frame.
                        \LState Set \textit{RIS\_Link\_Timer} according to duration.
                        \If{$\mathbf{Y}$'s \textit{RIS\_Link\_Timer} expires}
                            \LState Remove the RIS link.
                        \EndIf
                    \Else
                        \LState Abort current operation.
                        \LState Enter backoff state, then send R-RTS again.
                    \EndIf
                \EndIf
            \Else
                \LState $\mathbf{Y}$ replies R-CTS to source node.
                \LState $\mathbf{Y}$ waits for a SIFS time for DATA.
                \If{$\mathbf{Y}$ received DATA from the source node}
                    \LState Reply ACK to the previous hop node.
                    \LState Forward data to next hop as sending node $\mathbf{X}$.
                \Else
                    \LState Abort current operation. 
                    \LState Enter backoff state, then send R-CTS again.
                \EndIf
            \EndIf
        \EndIf
    \end{algorithmic}

\end{breakablealgorithm}

The RIS-DCF protocol for receiving nodes processes incoming R-RTS frames by accurately determining whether the node is the final destination or a relay node and then adjusting the RIS module phase accordingly to facilitate direct communication. It also handles collision detection, data reception, and error recovery, ensuring seamless data transfer even in network disturbances. The protocol also applies a retransmission mechanism, which ensures reliable communication by allowing nodes to retransmit lost or corrupted packets.

We have integrated the standard RTS/CTS protocol into the RIS-DCF protocol. The RTS/CTS scheme will be employed in two scenarios. First, direct one-hop LOS communication is enabled without needing relay nodes when the receiver node is the destination node. Second, when the relay node cannot provide RIS-assisted relay services, node R will continue to use conventional channels for communication. The relay node makes the decision on whether the installed RIS can be used for channel relay functions. With the proposed RIS-DCF protocol, the RIS controller can be integrated with the UAV's original wireless communication module, enabling efficient multi-hop relay network operations. 

The next section provides a detailed description of Algorithm 1 and Algorithm 2, focusing on the interaction between the sending and receiving nodes, including the definition and application of the RIS-assisted link and \textit{RIS\_Link\_Timer} mechanisms for relay nodes. Additionally, a comparison between the RIS-assisted link and pure conventional link scenarios is provided to illustrate the protocol's operation in different network environments.

\subsection{Detailed Illustration of the RIS-DCF Protocol}
This subsection provides a detailed explanation of the RIS-DCF protocol, covering the negotiation and transmission processes in both RIS-assisted link and pure conventional link scenarios.

\begin{figure*}[!t]
  \centering
  \includegraphics[width=6 in]{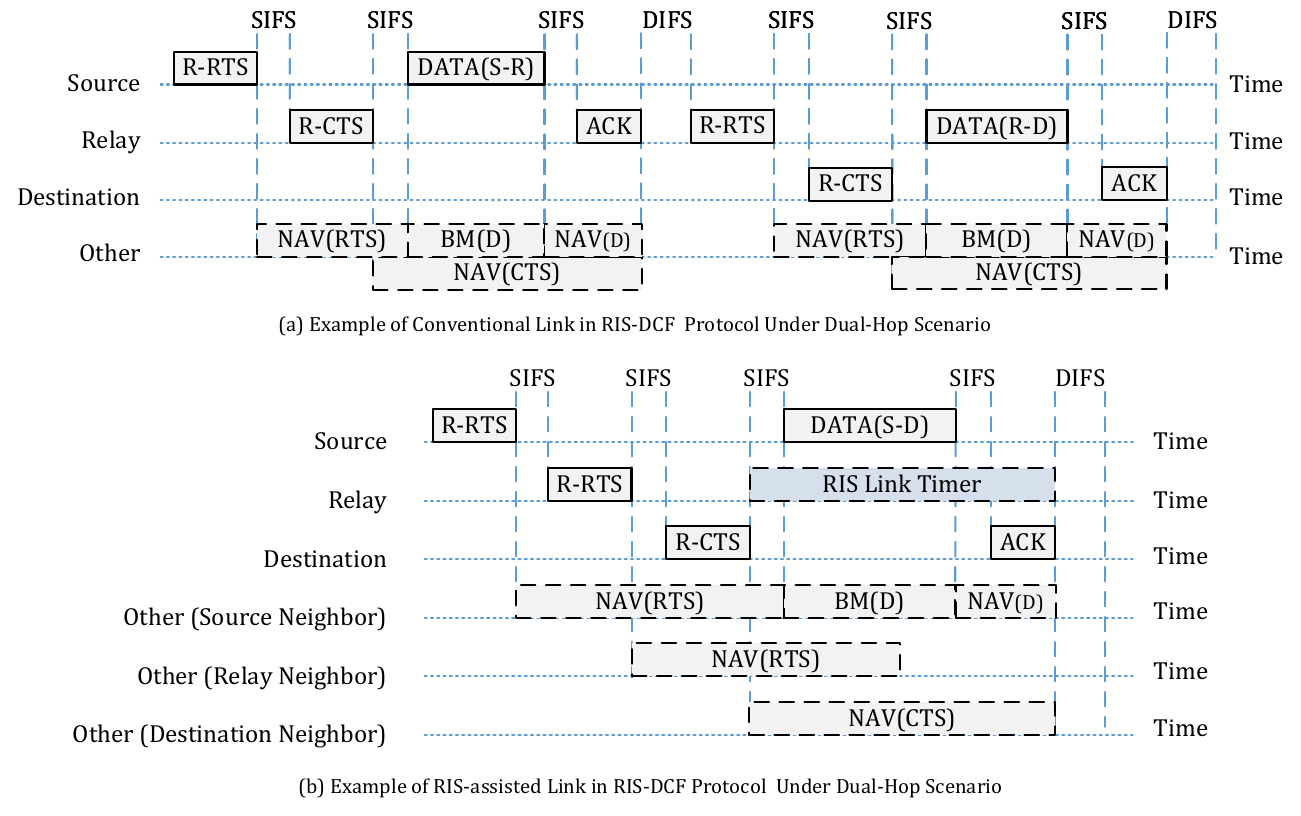}
  \caption{The RIS-assisted and conventional link cases in our proposed RIS-DCF protocol.}
  \label{fig_3}
\end{figure*}

As shown in Fig.~3(a), node S transmits data packets to node D without RIS assistance. First, node S determines the next-hop node address for node D and includes it in the R-RTS frame. When node S detects that the channel is idle and its backoff counter reaches zero, it sends the R-RTS to its next hop, node R. Upon receiving the R-RTS, relay node R checks its RIS module's status to decide whether to use RIS module. If RIS module is not used, node R waits for one SIFS period before sending an R-CTS to node S. After the SIFS, node S sends the data packet to node R. Upon receiving the packet, node R acknowledges receipt by sending an ACK to node S. Since node R is not the final destination, it forwards the R-CTS to the destination node, D. Node R then acts as the transmitting node, replicating the process node S used to send data packets to node D.

Fig.~3(b) illustrates the process through which node S transmits data packets to node D using a RIS-assisted channel. The communication process is explained step by step as follows:

When node S detects that the channel is clear and its backoff counter reaches zero, it sends an R-RTS to the next hop node, node R. Upon receiving the R-RTS, node R checks the status of its RIS module and adjusts the phase-shift matrix, establishing a RIS-assisted communication link between node S and node D. Node R then waits for a SIFS period before forwarding the R-RTS to node D.

When node D receives the R-RTS, it waits for another SIFS period before replying with an R-CTS, which is constructed based on the received R-RTS. At this point, node D can communicate with node S through the RIS-assisted channel, but it still requires the R-CTS to assess the RIS channel status with node R.

Once node S receives the R-CTS, it sends the data packets directly to node D, while node R listens passively. Node R extracts the duration of the R-CTS from the header and sets the RIS link duration timer, \textit{RIS\_Link\_Timer}, which controls the establishment time of the RIS-assisted channel.

After the transmission is completed, node D sends ACK packets, which node S acknowledges upon receipt. When the \textit{RIS\_Link\_Timer} expires, node R releases the RIS-assisted channel and prepares for the next communication session.

\subsection{Cooperation of Overhearing Nodes}

The previous section introduced the RIS-DCF protocol algorithms for receivers and transmitters of R-RTS and R-CTS. However, to make the protocol fully operational, it is necessary to collaborate with other nodes in the network to improve communication efficiency and prevent hidden node problems, making the overall communication protocol complete. The 802.11 standard introduces a virtual carrier sensing (VCS) scheme to achieve this operation. In this section, based on the improved 802.11 VCS scheme for self-organizing networks in \cite{zhuRDCFRelayenabledMedium2006}, we combine it with the unique feature of RIS-assisted relay, which only performs RIS reflection and does not forward through its original communication channel. This combination makes the RIS-DCF protocol more available.

In the RIS-DCF VCS scheme, the network allocation vector (NAV) time set by R-RTS grouping is classified into two types: the same or different destination address and receiver address. The NAV time is specified by the sending node in the Duration field of the R-RTS/R-CTS header, as shown in Fig.~2. The detailed duration information is shown in Table I. Here, R-RTS and R-CTS represent the duration of the corresponding frame, while SIFS represents the time interval between short frame spaces. The other timeline in Fig.~3 shows in detail the carrier detection and virtual carrier detection of other nodes with and without RIS assisted, where Busy Medium (BM) represents the busy channel detected by the carrier, and NAV ($\cdot$) represents VCS time caused by the packet ($\cdot$).

\begin{table}[h]
    \begin{center}
      \caption{ The Duration to NAV in RIS-DCF }
      \begin{tabular}{l|l} % <-- Alignments: 1st column center, 2nd center, and 3rd left, with vertical lines in between
        \hline
        \textbf{Packet Type} & \textbf{Duration}\\
        \hline
 R-RTS(same address) & $R\mbox{-}RTS + SIFS$\\
 R-RTS(different address) & $R\mbox{-}RTS + R\mbox{-}CTS + 2 SIFS$\\
 R-CTS(same address) & $DATA + 2SIFS$\\
 DATA & $ACK + SIFS$\\
        \hline
      \end{tabular}
    \end{center}
\end{table}

In the VCS scheme of the RIS-DCF protocol, there are two special situations to consider:

First, when the Relay node rejects RIS-assisted channels, neighboring nodes of the Source node face specific challenges. As shown in Fig.~3(a), when the Relay node decides not to use RIS assistance, the NAV duration set by the Source node can cause overlapping NAV and Busy Medium (BM) states for neighboring nodes during DATA frame transmission. In such cases, neighboring nodes of the Source node adjust their NAV timer values by receiving the Duration field in the DATA frame header, setting $\mathrm{NAV=\min \{Data\ Duration, NAV\}}$ to coordinate channel usage effectively.

\begin{figure}[h]
\centering
\includegraphics[width=3.4in]{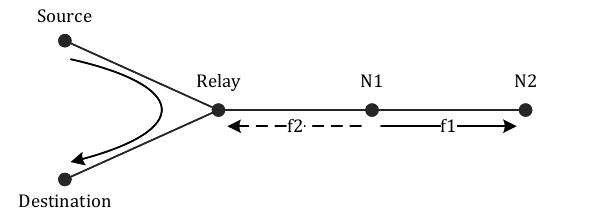}
\caption{The virtual carrier sensing of a neighboring node of the relay node.}
\label{fig_4b hidden terminal}
\end{figure}

Second, when the Relay node is in the RIS-assisted state, neighboring nodes of the Relay node encounter a different set of challenges. As illustrated in Fig.~\ref{fig_4b hidden terminal}, this scenario involves neighboring nodes of the Relay node when RIS assistance is used. The Relay node only controls R-RTS/R-CTS negotiation and RIS module without transmitting packets through the communication node. The neighboring node of the Relay node (N1) can communicate with other nodes (N2) via the f2 channel after the NAV (RTS) period, whereas the N1 attempting to send RTS packets to the Relay node through the f1 channel results in a timeout.

\section{Perfomance Analysis}
In this section, we establish a dual-hop access model based on the random access process model \cite{bianchiPerformanceAnalysisIEEE2000} to analyze the performance of the proposed RIS-DCF protocol. We compare the performance of the network with and without RIS assistance. Then, we extend the dual-hop access model to a multi-hop scenario to evaluate the performance of the RIS-DCF protocol in multi-hop relay networks. 

\subsection{Dual-hop Access Model}

\begin{figure}[h]
    \centering
    \includegraphics[width=3.3in]{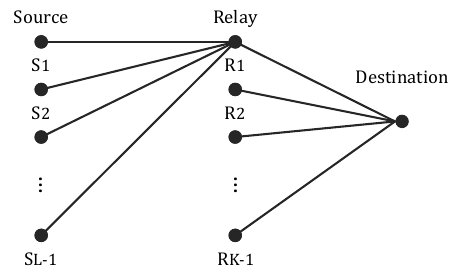}
    \caption{The dual-hop access model for RIS-DCF protocol.}
    \label{fig_4 dual-hop access}
\end{figure}

As shown in Fig.~\ref{fig_4 dual-hop access}, we consider a dual-hop access model via relay nodes. The model is primarily composed of three types of nodes: source node S, relay node R, and destination node D. At the relay node R, there are \(L\) neighboring nodes, including the source node S, which send packets to node D with an equal probability \(p\). Similarly, at the destination node D, there are $K$ neighboring nodes, including the relay node R, which also send packets to node D with an equal probability $p$. This dual-hop model employs a two-times \(p\)-persistent CSMA scheme for both channel reservation and RIS module reservation.

In RIS-assisted wireless networks, the time required for data transmission through the RIS channel can be calculated using the RIS-assisted channel efficiency $\eta$ defined in Eq.~(4):
\begin{equation}
T_\mathrm{RIS} = \frac{T_\mathrm{data}}{\eta}
\tag{8}
\end{equation}
where $T_\mathrm{RIS}$ represents the transmission time through the RIS channel, and $T_\mathrm{data}$ denotes the transmission time required for the same data payload of size $E$ bits over a conventional channel at the reference transmission rate.

In our model, successfully establishing RIS channels may encounter two collisions in different slots. The first collision occurs when the source node initiates sending messages to be sent to the relay node one hop away, and the neighboring nodes of the relay node send an R-RTS competing channel. The second collision occurs after the first hop connection is established successfully. When the relay node decides to forward R-RTS, $K$ nodes will compete for the corresponding channel.

If we denote $T_\mathrm{S}^\mathrm{R}$, $T_\mathrm{C1}^\mathrm{R}$ and $T_\mathrm{C2}^\mathrm{R}$, as the total time spent successfully transmitting via RIS, the time spent when the conflict occurred during the first R-RTS, when the conflict occurs during the second R-RTS period, respectively, we get
\begin{equation}
    \label{deqn_time}
    \left\{ 
        \begin{array}{ll}
 T_\mathrm{S}^\mathrm{R}=&R\mbox{-}RTS+SIFS+R\mbox{-}RTS+SIFS\\
                & + R\mbox{-}CTS+ SIFS+ H +T_{RIS} \\
                & +SIFS+ACK+DIFS, \\ 
 T_\mathrm{C1}^\mathrm{R}=&R\mbox{-}RTS+DIFS,\\ 
 T_\mathrm{C2}^\mathrm{R}=&R\mbox{-}RTS+SIFS+R\mbox{-}RTS+DIFS,\\
        \end{array}
    \right.
\end{equation}
where $ACK$ is the length of an acknowledgment (ACK) frame and $DIFS$ is the time interval of distributed coordination function inter-frame space (DIFS).

And then, we can also deduce the idle probability of the subchannel, the successful transmission probability of the first R-RTS, the collision probability of the first R-RTS, the successful transmission probability of the second R-RTS and the collision probability of the second R-RTS, they are represented by  $P_\mathrm{I}$, $P_\mathrm{S 1}$, $P_\mathrm{C 1}$, $P_\mathrm{S 2}$ and $P_\mathrm{C2}$ as follows:

\begin{equation}
    \label{deqn_prob}
    \left\{  
        \begin{array}{ll}  
 P_\mathrm{I}&=(1-p)^L, \\  
 P_\mathrm{S1}&= Lp(1-p)^{L-1},\\  
 P_\mathrm{C1}&=1-Lp(1-p)^{L-1}-(1-p)^L,\\
 P_\mathrm{S2}&= (1-p)^{K-1},\\  
 P_\mathrm{C2}&=1-(1-p)^{K-1},\\
        \end{array}  
    \right. 
\end{equation}
where $p$ is the transmission probability. Here, we assume every node in this model remains the same transmission probability in different slots. 

Next, we can calculate the end-to-end system saturation throughput with RIS assistance, denoted as $S_{R}$, as follows:

\begin{equation}
    \label{deqn_10}
 S_{R}= 
    \frac{P_{\mathrm{S1}} P_{\mathrm{S2}} E}
    {P_{\mathrm{I}}\sigma+
 P_{\mathrm{C1}} T_{\mathrm{C1}}^\mathrm{R}+
 P_{\mathrm{S1}}P_{\mathrm{S2}}T_{\mathrm{S}}^\mathrm{R}+
 P_{\mathrm{S1}}P_{\mathrm{C2}} T_{\mathrm{C2}}^\mathrm{R}}
\end{equation}
where $\sigma$ is the duration of an empty slot time.

The performance analysis for the conventional link under our dual-hop access model follows a similar derivation, detailed in the Appendix~A. The end-to-end system saturation throughput without RIS assistance can be expressed as:
\begin{equation}
    \label{deqn_11}
 S_{C}= 
    \frac{P_{\mathrm{S1}} P_{\mathrm{S2}} E}
    {P_{\mathrm{I}}\sigma+
 P_{\mathrm{C1}} T_{\mathrm{C1}}^\mathrm{C}+
 P_{\mathrm{S1}}P_{\mathrm{S2}}T_{\mathrm{S}}^\mathrm{C}+
 P_{\mathrm{S1}}P_{\mathrm{C2}} T_{\mathrm{C2}}^\mathrm{C}}
\end{equation}

The throughput gain, denoted as $\kappa$, is defined as the ratio of the saturation end-to-end throughput with RIS assistance to that without assessing the overall benefit of the proposed protocol and model.
\begin{equation}
\kappa = S_{R}/S_{C}
\end{equation}

By utilizing the throughput gain as $\kappa$ and saturation throughput $S_{R}$ and $S_{C}$, we can evaluate the performance of the RIS-DCF protocol for RIS nodes within a dual-hop access model.

\subsection{Performance of Multi-hop Senerio}

To extend the dual-hop access model to an $\mathrm{m}$-hop scenario, we classify the time and probability parameters. Building upon the characteristics of the dual-hop and conventional single-hop channel models, we extend these results to an $\mathrm{m}$-hop scenario.

Assuming proper functionality of all RIS links, we analyze the successful transmission and collision times for the $\mathrm{m}$-th hop as follows. The total time spent successfully transmitting through the $\mathrm{m}$-hop via RIS links is denoted by $T_{S\mathrm{m}}^{R}$:

\begin{equation}
T_{S\mathrm{m}}^{R} = 
\begin{cases}
\frac{\mathrm{m}}{2} T_{S}^{R}, & \text{if } \mathrm{m} \text{ is even} \\
\frac{\mathrm{m}-1}{2} T_{S}^{R} + T_{S}^{C}, & \text{if } \mathrm{m} \text{ is odd}
\end{cases}
\end{equation}

This implies that when $\mathrm{m}$ is even, all hops can be connected via RIS channels, while for odd $\mathrm{m}$, at least one hop must use a conventional channel.

Next, we analyze the time spent when a conflict occurs during the $i$-th hop, denoted by $T_{Ci}^{R}$:  

\begin{equation}
T_{Ci}^{R} = 
\begin{cases}
T_{C2}^{R}, & \text{if } i \text{ is even} \\
T_{C1}^{R}, & \text{if } i \text{ is odd}
\end{cases}
\end{equation}

The time analysis for the conventional link in a multi-hop scenario follows a similar derivation, which is detailed in the Appendix~B.

To accurately evaluate the transmission process, we introduce the probabilities of collision and successful transmission:

\begin{equation}
P_{\mathrm{C}i} = \begin{cases}
P_{\mathrm{C}1}, & \text{if } i=1 \\
P_{\mathrm{C}2}, & \text{if } i>1
\end{cases} , \quad 
P_{\mathrm{S}i} = \begin{cases}
P_{\mathrm{S}1}, & \text{if } i=1 \\
P_{\mathrm{S}2}, & \text{if } i>1
\end{cases} 
\end{equation}

The following derivation applies to both RIS and conventional channels. Since a node participates in transmission starting from the second hop onward, the channel remains active between subsequent hops, similar to the second hop in the dual-hop access model. The total transmission time for the multi-hop scenario, denoted by \( T_{\text{total}} \), is calculated as:

\begin{equation}
T_{\text{total}} = P_{\mathrm{C1}} T_{\mathrm{C1}} + 
\sum_{j=2}^\mathrm{m} \left(\prod_{i=1}^{j-1} P_{\mathrm{S} i}\right) P_{\mathrm{C} j} T_{\mathrm{C}j} + 
\prod_{i=1}^m P_{\mathrm{S} i} T_{\mathrm{S}}
\end{equation}

Based on this model, we define the $\mathrm{m}$-hop end-to-end system saturation throughput $S_\mathrm{m}$ as follows:

\begin{equation}
S_\mathrm{m} = \frac{\prod_{i \in [1, \mathrm{m}]} P_{Si} E}  
{P_{\mathrm{I}} \sigma + T_{\text{total}}}
\end{equation}

This formula reflects the impact of transmission success probability and collision time on the $\mathrm{m}$-hop saturation throughput. System performance can be optimized by properly adjusting channel parameters.

\section{Perfomance Evaluation}

This section evaluates our proposed network model and the RIS-DCF protocol in a dual-hop access model using numerical results. First, we compare the impact of throughput gain as a function of SNR with varying numbers of RIS elements. Second, we compare the saturation throughput in RIS-assisted links and conventional links, varying the number of source nodes and relay nodes. Third, we evaluate the effects of packet length, initial backoff window length, and transmission probability on saturation throughput with and without RIS assistance.
The parameter settings in the simulation are summarized in Table II unless otherwise specified or varied.

\begin{table}[h]
    \begin{center}
      \caption{Simulation Parameters}
      \begin{tabular}{c|c|l} % <-- Alignments: 1st column center, 2nd center, and 3rd left, with vertical lines in between
        \hline
        \textbf{Parameter} & \textbf{Value} & \textbf{Definition}\\
        \hline
        $f$ & 5.8 GHz & Frequency of communication\\
        $d$ & 1000 m & Node-to-node distance\\
        $m$ & 2.5 & Nakagami-m parameter\\
        $N_0$ & -85 dBm  & Average noise power \\
        $P_s$ & 0 dBm & Average transmit power\\
        \hline
        $\mathrm{R_{conv}}$ & 1 Mbps & Base rate for conventional channel\\
 DIFS & 128 $\mu$s & DIFS (DCF Interframe Space) Time\\
 SIFS & 28 $\mu$s & SIFS (Short Interframe Space) Time\\
 Slot Time & 50 $\mu$s & Slot time duration\\
        \hline
 R-RTS & 208 bits & R-RTS frame length\\
 R-CTS & 160 bits & R-CTS frame length\\
 ACK & 240 bits & ACK frame length\\
 PHY Headers & 128 bits & Physical layer header length\\
 MAC Headers & 272 bits & Multiple access layer header length\\
 Packet Payload & 1000 bytes & Length of packet payload\\
        \hline
        $L$ & 5 & Number of Source node\\
        $K$ & 6 & Number of Relay node\\
        $\mathrm{m}$&2&Number of hops\\
        $n$ & 3 & Maximum backoff stage\\
        $W$ & 32 & Initial size of the backoff window\\
        \hline
      \end{tabular}
    \end{center}
\end{table}

\begin{figure}[h]
    \centering
    \includegraphics[width=3.2in]{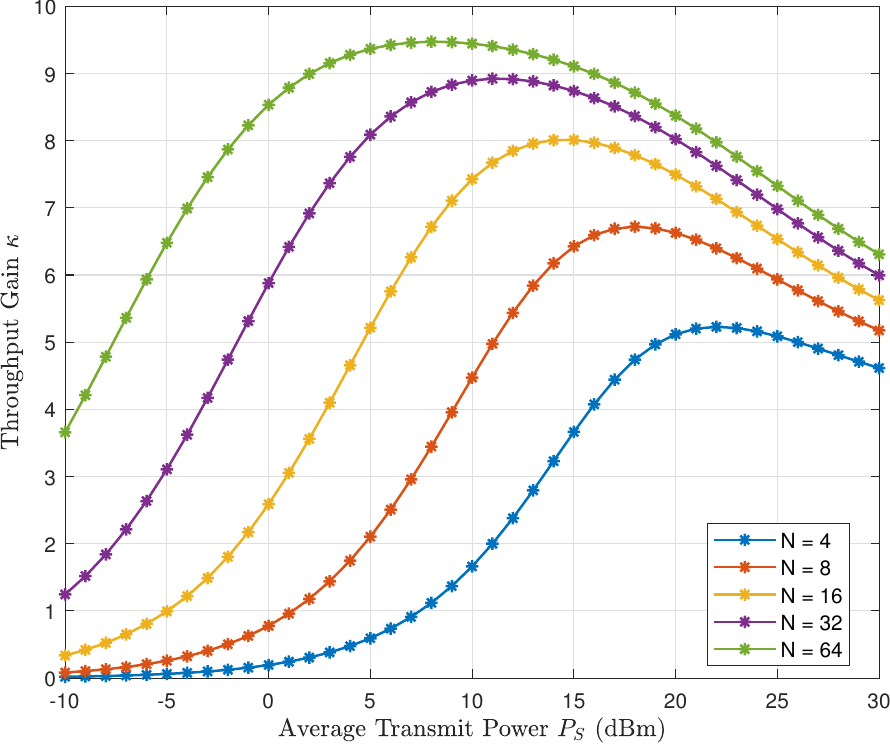}
    \caption{The relationship between throughput gain $\kappa$ and average transmit power $P_s$}
    \label{fig_4}
\end{figure}

Fig.~\ref{fig_4} depicts the throughput gain as a function of the average transmit power of transmitting nodes $P_S$ with varying numbers of RIS elements $N$. The transmission power will affect the received SNR. As the SNR gradually increases, the throughput gain of all RIS will reach a maximum value and then decrease, gradually stabilizing. The employment of RIS channels imposes a requirement on the SNR. When the SNR is very low, the link gain obtained from the RIS-assisted link is significantly smaller than that of the conventional link due to large path loss. 
Under high SNR, the throughput gain of RIS is substantially enhanced, leading to significantly higher throughput compared to conventional links under varying conditions. To evaluate the performance of RIS-assisted and conventional links under extreme conditions, we analyze the throughput gain of RIS elements at $N = 4$, $8$, and $16$, with a transmission power of 0 dBm. In this scenario, the low transmit power results in a reduced SNR, highlighting channel effects and enabling the analysis of other influencing parameters.

Additionally, Fig.~\ref{fig_4} demonstrates the diversity gain associated with the number of RIS elements, implying that the end-to-end throughput correspondingly improves as the number of RIS elements increases. The size of the RIS module is directly proportional to $N$ and the wavelength $\lambda$. As $N$ increases, the RIS module size may become huge, potentially hindering the mobility of UAVs and other equipment carrying the module \cite{sihlbomReconfigurableIntelligentSurfaces2023}. Therefore, to ensure the mobility of UAVs, the maximum number of RIS elements used here is limited to 64. 

\begin{figure}[t]
    \centering
    \includegraphics[width=3.2in]{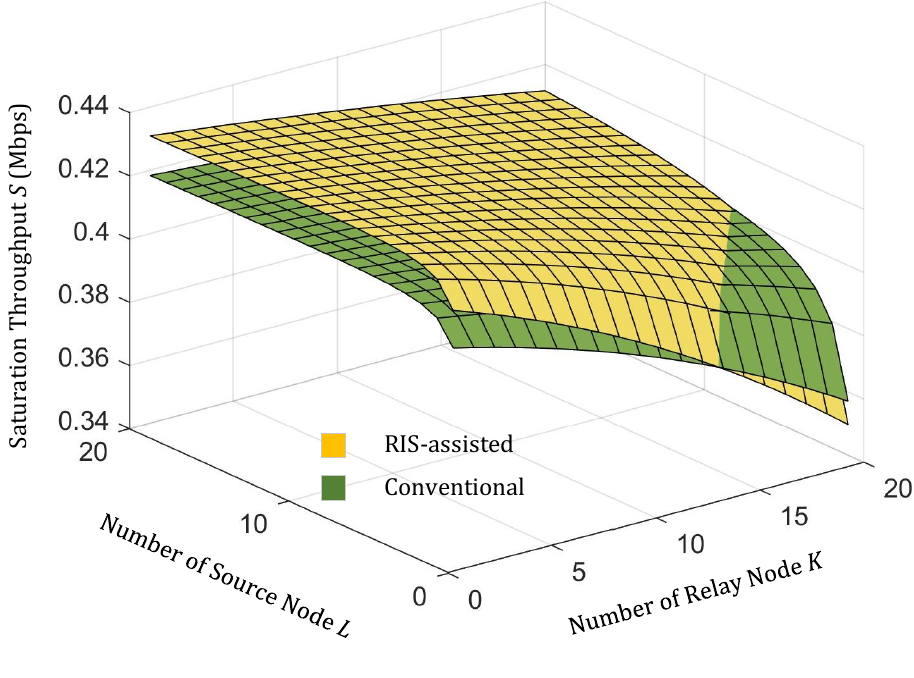}
    \caption{The comparison of saturation throughput with and without RIS assistance with varying numbers of source nodes $L$ and relay nodes $K$.}
    \label{fig_5a}
\end{figure}

\begin{figure}[t]
    \centering
    \subfloat[]{\includegraphics[width=3.2in]{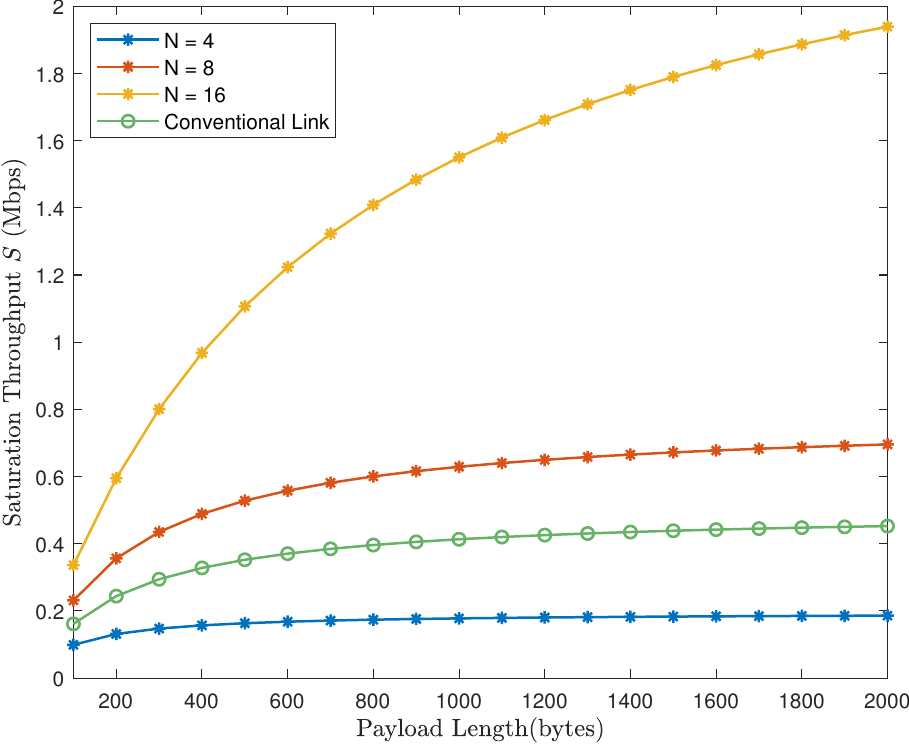}%
    \label{fig_second_case}}
    \hfil
    \subfloat[]{\includegraphics[width=3.2in]{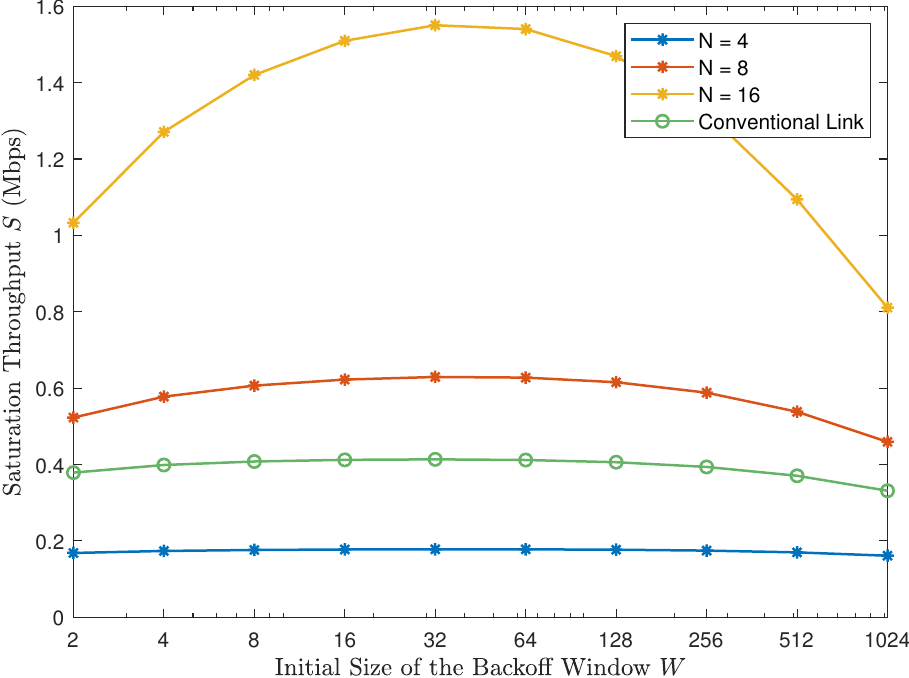}%
    \label{fig_3_case}}
    \caption{The saturation throughput of RIS-DCF: (a) impact of payload length, (b) impact of backoff window.}
    \label{fig_6}
\end{figure}

In Fig.~\ref{fig_5a}, the RIS efficiency $\eta$ is set to 0.5, which is a pivotal value for evaluating the performance of the RIS-DCF protocol with and without RIS assistance. Fig.~\ref{fig_5a} depicts the relationship between the number of competing nodes in the first and second hops and the saturation throughput. The yellow surface represents the throughput achieved with the RIS-assisted link, whereas the green surface denotes the throughput with the conventional link. The RIS-assisted link significantly outperforms the conventional link in terms of saturation throughput when $L$ and $K$ are small. However, as the number of nodes competing for access increases, the throughput of RIS-assisted links decreases more rapidly than that of conventional channels, ultimately falling below the throughput of conventional channels.

In the dual-hop access model, interference from neighboring nodes at the receiver emerges as the main factor reducing throughput performance. The reduction in throughput is primarily attributed to the operation of the second hop node, which experiences pure collisions without involving data transmission. These collisions significantly impact the network's overall throughput model. Due to the higher cost associated with the second hop in the RIS channel, this hop is more susceptible to collisions, resulting in a higher saturation throughput in the conventional link compared to the RIS-assisted state.

\begin{figure}[t]
    \centering
    \includegraphics[width=3.2in]{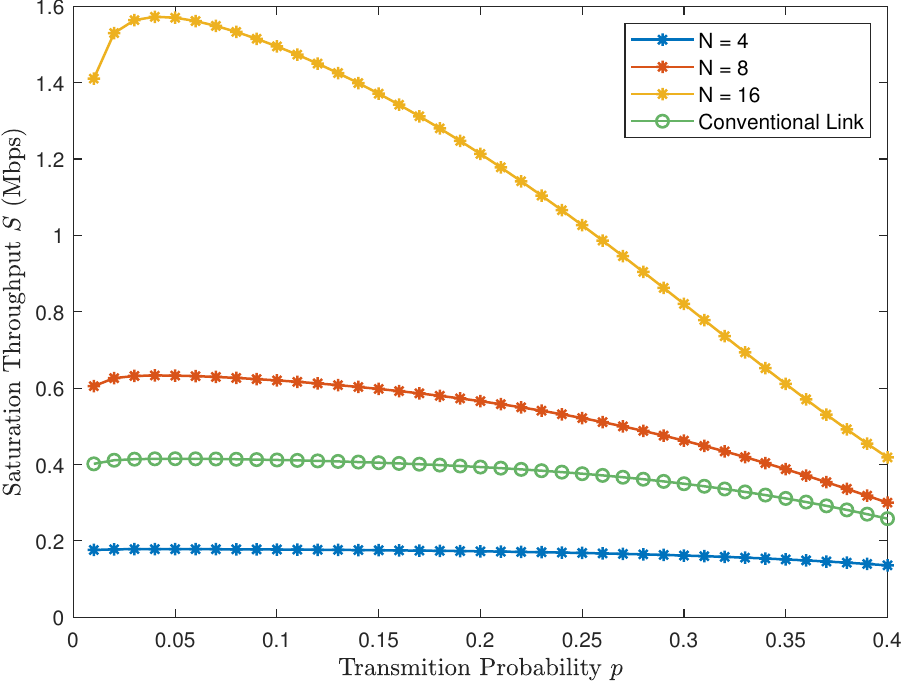}
    \caption{The impact of transmission probability on saturation throughput.}
    \label{fig_5b}
\end{figure}

Fig.~\ref{fig_6}(a) depicts the throughput variation with packet load, revealing the effectiveness of RIS-assisted link processing for both long and short packets. The increase in the number of RIS elements $N$ consistently results in higher throughput, particularly when utilizing long packets. 
However, as the packet length increases, the throughput increases and gradually converges to a specific value, particularly in scenarios with numerous RIS elements and high diversity gain, where an increase in RIS elements leads to faster growth and later convergence.
This underscores that the proposed model is more suitable for establishing scenarios with internal relay backbone links in the network.

Fig.~\ref{fig_6}(b) depicts the throughput variation with the initial size of the backoff window. The throughput initially increases with the size of the initial backoff window, reaching a peak at a window size of 32, after which it declines. The optimal backoff window size corresponds to the point where throughput is maximized, balancing the tradeoff between collision probability and delay. Configurations with more RIS elements ($N = 8$ and $N = 16$) exhibit more pronounced variations in throughput with changes in the backoff window size, with both the rise and fall being steeper than the conventional link and the $N = 4$ configuration. This sensitivity suggests that higher transmission rates achieved with more RIS elements result in significant throughput fluctuations with varying backoff window sizes.

Fig.~\ref{fig_5b} depicts the throughput variation with transmission probability. As the transmission probability increases, the throughput initially reaches a maximum and then begins to decrease. Eventually, the performance of the RIS-DCF protocol with a RIS-assisted link approaches that of the conventional link. These results suggest that the proposed protocol is only suitable for scenarios with a low number of neighboring nodes and for establishing internal relay backbone links within the network. Conversely, it is unsuitable for crowded and dense networking environments. Therefore, adaptive switching can be considered to address the limitations. By assessing the RIS channel quality from channel information, RIS can be utilized when the channel quality is good, whereas conventional channel can be employed when the quality is poor, as proposed in the RIS-DCF protocol.

\begin{figure}[t]
    \centering
    \includegraphics[width=3.2in]{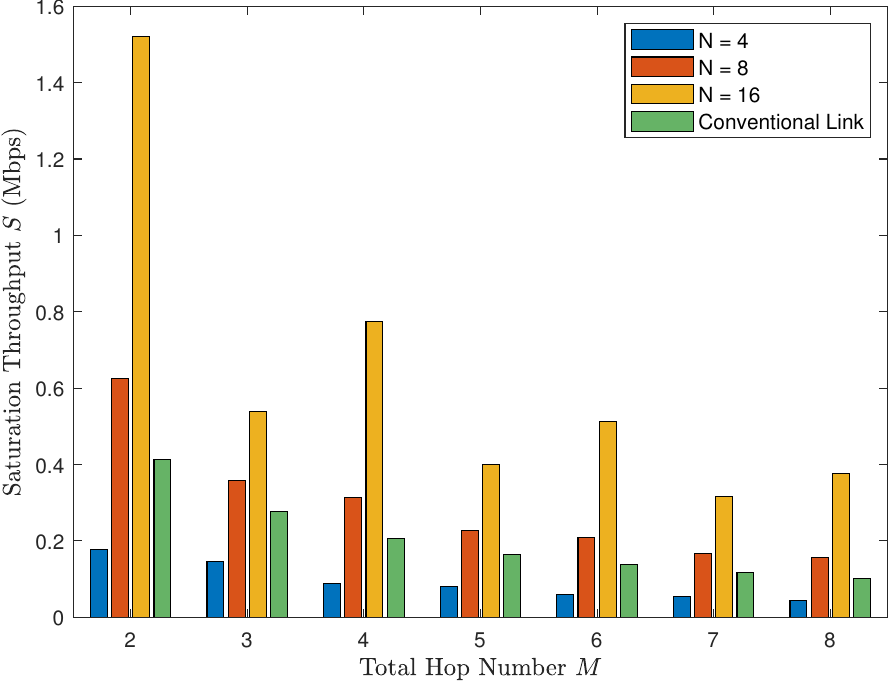}
    \caption{The impact of relay hop number $M$ on saturation throughput.}
    \label{fig_hops}
\end{figure}

Fig.~\ref{fig_hops} depicts the throughput variation with the number of relay hops. As the number of relay hops increases, throughput decreases. This decrease can be attributed to the increased number of relay hops, leading to more cost and more collisions. 
When the number of relay hops is small, the throughput gain provided by RIS is substantial. However, as the number of hops increases, the collision overhead in the second hop of the two-hop model becomes more pronounced, which reduces the overall throughput benefit of RIS. Nonetheless, the diversity gains still ensure that the performance of RIS-assisted links remains superior to that of conventional links when $N = 8$ and $N = 16$.
Since we introduced the multi-hop scheme as a cascade of double and single hops, the throughput of the multi-hop RIS link in $N = 16$ damping oscillates between odd and even hops, eventually leading to convergence. From this perspective, a higher number of RIS elements mitigates the throughput decrease caused by the increased number of relay hops.

Notably, the case with $N=4$ represents a distinct scenario in Figs. \ref{fig_6}, \ref{fig_5b} and \ref{fig_hops}. Here, the received SNR is notably low, because the detrimental effects of path loss caused by long distance overshadow the diversity gain provided by the RIS. Consequently, the throughput achieved in this scenario is lower than that observed in standard communication channels. This observation underscores the critical importance of relay-based evaluation of channel conditions for effectively deploying RIS modules.

\section{Conclusion}

In this paper, we proposed a RIS-assisted MANET that integrated the RIS modules with the existing communication system of nodes.
By fully leveraging the seamless connectivity and diversity gain facilitated by RIS, this protocol significantly enhances the performance of MANET while effectively addressing hidden station issues.
Our comprehensive analysis and simulations illustrate notable throughput enhancements achieved through the incorporation of RIS components and the enhancement of SNR. Additionally, RIS-DCF shows improved performance with longer packets and exhibited sensitivity to parameters such as backoff window size and transmission probability. Furthermore, numerical results reveal that RIS-assisted links are more suitable for sparse and less competitive wireless multi-hop relay networks.
While the increased deployment of RIS elements is correlated with higher throughput, it is essential to consider mobility constraints when determining the optimal number of RIS elements. 
In summary, our study highlights the potential of RIS-assisted MANETs with the integration of node communication equipment, as demonstrated by the performance enhancements achieved with RIS-DCF.

\bibliographystyle{IEEEtran}
\bibliography{240709RSC_arxiv.bib}

\appendices

\section{The Saturation End-to-End Throughput without RIS-Assisted}

If we denote $T_\mathrm{S}^\mathrm{C}$, $T_\mathrm{C1}^\mathrm{C}$ and $T_\mathrm{C2}^\mathrm{C}$, as the total time spent successfully transmitting without RIS-assisted, the time spent when the conflict occurred during the first R-RTS without RIS-assisted, when the conflict occurred during the second R-RTS period without RIS-assisted, respectively, we get
\begin{equation*}
    \left\{ 
        \begin{array}{ll}
            T_\mathrm{S}^\mathrm{C}=&2(R\mbox{-}RTS+SIFS + R\mbox{-}CTS+ SIFS\\
                &+ H +T_{data} +SIFS+ACK+DIFS,) \\ 
            T_\mathrm{C1}^\mathrm{C}=&R\mbox{-}RTS+DIFS,\\ 
            T_\mathrm{C2}^\mathrm{C}=&R\mbox{-}RTS+DIFS,\\
        \end{array}
    \right.
\end{equation*}

And then, under the same dual-hop access model as the RIS-assisted scenario,  $P_\mathrm{I}$, $P_\mathrm{S 1}$, $P_\mathrm{C 1}$, $P_\mathrm{S 2}$ and $P_\mathrm{C2}$ maintain the same in the conventional scenario as Eq. (10). And we can derive the saturation end-to-end system throughput without RIS-assisted $S_\mathrm{C}$ Eq. (12) in performance analysis.
    
\section{The $\mathrm{m}$-Hop Time Analysis without RIS-Assisted}

Due to the absence of RIS relays in dual-hop transmission, the time analysis of the $\mathrm{m}$-hop transmission without RIS-assisted is simplified. The total time spent successfully transmitting through the $\mathrm{m}$-hop via conventional links only is denoted by $T_{S\mathrm{m}}^{C}$:

\begin{equation*}
  T_{S\mathrm{m}}^\mathrm{C} = \frac{T_\mathrm{S}^\mathrm{C}}{2}
\end{equation*}
  
From Appendix~A, it can be concluded that $T_\mathrm{C1}^\mathrm{C}=T_\mathrm{C2}^\mathrm{C}$, and the time spent when a conflict occurred during the $i$-th hop via conventional links only, denoted by $T_{Ci}^{R}$, is given by:  
\begin{equation*}
  T_{Ci}^{C} = T_\mathrm{C}^\mathrm{C} = R\mbox{-}RTS+DIFS
\end{equation*}
where $T_\mathrm{C}^\mathrm{C} = T_\mathrm{C1}^\mathrm{C}=T_\mathrm{C2}^\mathrm{C}$ denotes the total time spent when a conflict occurred via conventional links only in the dual-hop access model.

\vfill
\end{document}